\documentclass[11pt,a4paper]{article}
\usepackage[T1]{fontenc}
\usepackage[utf8]{inputenc}
\usepackage{lmodern}
\usepackage[margin=1in,includefoot,footskip=20pt]{geometry}
\usepackage{amsmath,amssymb,graphicx,array,booktabs}
\usepackage[authoryear]{natbib}
\usepackage{xurl}
\usepackage[hidelinks,unicode]{hyperref}
\usepackage{microtype}
\DeclareUnicodeCharacter{2010}{-}
\newcommand{\arcsec}{\ensuremath{^{\prime\prime}}}
\newcommand{\arcmin}{\ensuremath{^{\prime}}}

\hypersetup{pdftitle={Adaptive Optics for the Extremely Large Telescope: overcoming Point Spread Function challenges with MORFEO},
 pdfauthor={Carmelo Arcidiacono, Elisa Portaluri, Marco Gullieuszik, Michele Cantiello, Francesca Annibali, Guido Agapito, Demetrio Magrin, Vincenzo Cianniello, Paolo Ciliegi}}
\begin{document}
\begin{center}
{\Large\bfseries Adaptive Optics for the Extremely Large Telescope: overcoming Point Spread Function challenges with MORFEO\par}
\vspace{9pt}
{\normalsize Carmelo Arcidiacono\textsuperscript{1,3,*}, Elisa Portaluri\textsuperscript{2,3}, Marco Gullieuszik\textsuperscript{1}\\[2pt]
Michele Cantiello\textsuperscript{2}, Francesca Annibali\textsuperscript{4}, Guido Agapito\textsuperscript{3,5}\\[2pt]
Demetrio Magrin\textsuperscript{1,3}, Vincenzo Cianniello\textsuperscript{6}, Paolo Ciliegi\textsuperscript{4}\par}
\vspace{7pt}
{\small
\textsuperscript{1}INAF Osservatorio Astronomico di Padova, Vicolo dell'Osservatorio 5, Padova, 35121, Italy\\[2pt]
\textsuperscript{2}INAF Osservatorio Astronomico d'Abruzzo, Vicolo M. Maggini snc, Teramo, 35121, Italy\\[2pt]
\textsuperscript{3}ADONI, Italy\\[2pt]
\textsuperscript{4}INAF Osservatorio di Astrofisica e Scienza dello spazio di Bologna, via Gobetti 93/3, Bologna, 40129, Italy\\[2pt]
\textsuperscript{5}INAF Osservatorio Astrofisico di Arcetri, Largo Fermi, 5, Firenze, 50125, Italy\\[2pt]
\textsuperscript{6}INAF Osservatorio Astronomico di Capodimonte, Salita Moiariello 16, Napoli, 80131, Italy\par}
\end{center}
\begingroup
\renewcommand{\thefootnote}{}
\footnotetext{\textsuperscript{*}Arcidiacono Carmelo. \href{mailto:carmelo.arcidiacono@inaf.it}{\nolinkurl{carmelo.arcidiacono@inaf.it}} (C. Arcidiacono); \href{mailto:elisa.portaluri@inaf.it}{\nolinkurl{elisa.portaluri@inaf.it}} (E. Portaluri). \url{https://www.oapd.inaf.it/} (C. Arcidiacono).\\
ORCID(s): \href{https://orcid.org/0000-0003-0142-8108}{0000-0003-0142-8108} (C. Arcidiacono); \href{https://orcid.org/0000-0002-1242-7242}{0000-0002-1242-7242} (E. Portaluri); \href{https://orcid.org/0000-0002-7296-9780}{0000-0002-7296-9780} (M. Gullieuszik); \href{https://orcid.org/0000-0003-2072-384X}{0000-0003-2072-384X} (M. Cantiello); \href{https://orcid.org/0000-0003-3758-4516}{0000-0003-3758-4516} (F. Annibali); \href{https://orcid.org/0000-0002-6382-2613}{0000-0002-6382-2613} (G. Agapito); \href{https://orcid.org/0000-0003-0312-313X}{0000-0003-0312-313X} (D. Magrin); \href{https://orcid.org/0000-0003-1111-3628}{0000-0003-1111-3628} (V. Cianniello); \href{https://orcid.org/0000-0002-0447-4620}{0000-0002-0447-4620} (P. Ciliegi)}
\endgroup

\begin{abstract}
MORFEO is the multi-conjugate adaptive optics (MCAO) relay for the Extremely Large Telescope (ELT) of the European Southern Observatory. It combines the adaptive mirrors of the telescope (M4-M5) with two post-focal deformable mirrors to deliver a stable, diffraction-limited point spread function over a field approaching 1\arcmin. MORFEO senses the atmosphere using six sodium laser guide stars and up to three natural guide stars distributed across a $\sim$160\arcsec\ technical field, enabling sky coverage levels of order 50\% at the South Galactic Pole. Through its two gravity-invariant ports, MORFEO feeds the MICADO first-light imager and spectrograph with a corrected field of 51\arcsec and HARMONI, the high-angular resolution integral field spectrograph.

SHARP, a concept for a multi-object near-infrared spectrograph at the ELT, adopts MORFEO as its reference MCAO architecture for evaluating achievable image quality and scientific performance. In this paper, we summarise the MCAO capabilities delivered by MORFEO, its optical quality, tomographic performance, stability of the corrected focal plane, and the accuracy of telemetry-based PSF reconstruction, and discuss their relevance for the development of SHARP and other wide-field spectroscopic concepts at the ELT.
\end{abstract}

\subsection*{Highlights}
\begin{itemize}
\setlength{\itemsep}{0pt}

\item MORFEO provides wide-field MCAO correction over a 120\arcsec\ field.
\item Delivers diffraction-limited imaging for MICADO at ELT scale.
\item Includes PSF reconstruction from AO telemetry for sparse fields.
\item Enables AO-corrected access to a second port for spectroscopy.
\end{itemize}

\section{Introduction}\label{sec:introduction}

Adaptive optics (AO) has transformed ground-based optical and near-infrared astronomy by restoring diffraction-limited performance on large telescopes and thereby enabling high-contrast, high-resolution imaging and spectroscopy. The Extremely Large Telescope (ELT) (\cite{gilmozzi_european_2007}) integrates AO at the system level through the adaptive quaternary mirror M4 and the fast tip--tilt mirror M5; together with the Laser Projection System (\cite{nijenhuis_development_2022}), these enable several AO modes ranging from Single-conjugate AO (SCAO) (\cite{babcock_possibility_1953}) and Ground Layer AO (GLAO) (\cite{rigaut_ground_2002}) to Laser Tomography AO (\cite{foy_feasibility_1985}) and Multi--Conjugate AO (\cite{beckers_increasing_1988}). Achieving stable diffraction-limited imaging at ELT aperture scales requires wide-field AO architectures capable of mitigating anisoplanatism and telescope-induced aberrations across arcminute fields.

MORFEO (Multi-conjugate adaptive Optics Relay For ELT Observations), formerly MAORY (\cite{diolaiti_maory_2016}), is the MCAO module of the ELT. It combines telescope adaptive optics with two post-focal deformable mirrors (DM) conjugated at high-altitude layers (nominally $\sim$6.5~km and $\sim$17.5~km). The AO system uses the ELT constellation of six sodium Laser Guide Stars (LGSs) for high-order sensing, along with up to three faint Natural Guide Stars (NGSs) for low-order and truth sensing. This architecture delivers an adaptive correction over a technical field of $\sim$160\arcsec and supports the AO needs of two gravity-invariant instrument ports: the MICADO (\cite{davies_micado_2010}) first-light imager/spectrograph and a second port assigned to the HARMONI spectrograph (\cite{thatte_e-elt_2016}). For MICADO, MORFEO optimises diffraction-limited performance over the field imaged by the camera, allowing precision photometry, astrometry, and morphology studies in a range of science cases (\cite{davies_micado_2021,rodeghiero_performance_2021,clenet_micado_2022}). The wider corrected field accessible from the second port, of order $\sim$2--2.5\arcmin, enables concepts such as multi-object spectroscopy at the ELT scale.

A central challenge for ELT-class AO is the time- and field-dependent variability of the point-spread function (PSF), driven by atmospheric turbulence, telescope-induced perturbations, and residual control errors. In high-latitude extragalactic fields, the scarcity of suitably bright stars complicates empirical PSF characterisation. MICADO therefore adopts a blind telemetry-driven PSF reconstruction strategy (\cite{veran_estimation_1997, wagner_point_2018,grazian_status_2022}), which uses AO telemetry and control matrices to recover the science PSF without relying on isolated point sources. MORFEO is designed to provide sky coverage of at least 50\% at high Galactic latitude, K-band Strehl ratios of $\gtrsim 30\%$ under median conditions, and the PSF stability required to minimise PSF-driven systematics in photometry and morphology (\cite{busoni_maorymorfeo_2022, ciliegi_maorymorfeo_2022}).

The SHARP project, a concept for a multi-mode near-infrared spectrograph for the ELT, uses MORFEO as its reference MCAO relay when assessing the delivered image quality and AO performance budgets (\cite{saracco_sharp_2022}). In this paper, we summarise the ELT AO context and the architecture of MORFEO (Sect.~\ref{sec:introduction}--\ref{sec:morfeo-arch}), discuss science-driven performance considerations (Sect.~\ref{sec:requirements}--\ref{sec:performance}), and outline the PSF-reconstruction strategy and its validation (Sect.~\ref{sec:psfr}). We also present observational tools (AETC and ScopeSim) adapted to the MORFEO-MICADO system (Sect.~\ref{sec:etc}).

\section{Adaptive Optics at the ELT}\label{sec:elt-ao}

The ELT integrates AO at the telescope level through its quaternary deformable mirror M4 (\cite{vernet_elt_2019}), conjugated to the ground layer, and the fast tip--tilt mirror M5 (\cite{vernet_elt_2021}). M4 (2.4\,m class, $>$5000 actuators, operated up to $\sim$1\,kHz) provides high-order wavefront correction for atmospheric turbulence and telescope perturbations, while M5 delivers fast image stabilisation before the light reaches the instruments (\cite{biasi_e-elt_2015, chiozzi_elt_2018}). Together, they form the AO core of the ELT’s five-mirror optical train and feed the Nasmyth platforms.

The observatory will deploy six sodium LGSs (\cite{nijenhuis_development_2022}), with provision for expansion to eight, to ensure high sky coverage across all AO modes. The lasers excite mesospheric sodium at 589.2\,nm and project a constellation of six artificial stars arranged on a 1.5\arcmin diameter asterism (\cite{ nijenhuis_development_2022}). The artificial laser sources are essential to access the full sky, providing the high-order wavefront measurements required by the AO system; the overall sky coverage is therefore limited by the availability of one or more NGSs for tip-tilt and low-order correction.

\begin{figure}[htbp]
  \centering
  \includegraphics[width=.88\linewidth,height=.34\textheight,keepaspectratio]{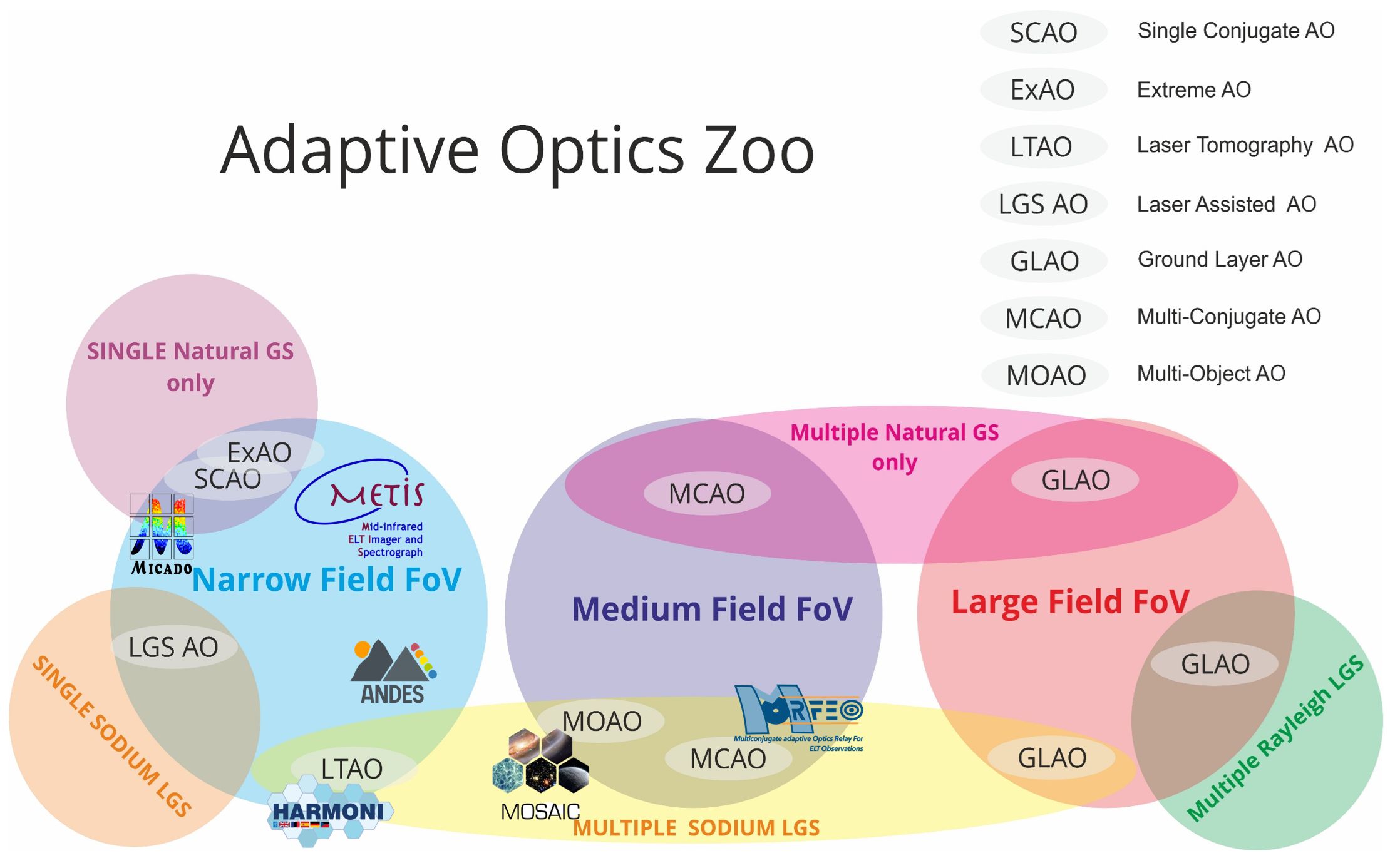}
  \caption{The ``Adaptive Optics Zoo’’ summarising the main AO modes employed on the ELT. The diagram highlights their typical field of view, dependence on natural or laser guide stars, and associations with ELT instruments, from narrow-field SCAO/ExAO to wide-field GLAO and multi-conjugate laser-assisted systems such as MORFEO.}
  \label{fig:adaptive_optics_zoo}
\end{figure}

The ELT AO architecture supports multiple modes: in SCAO, a single DM enables the highest Strehl ratio around the bright NGS but suffers rapid off-axis degradation. GLAO addresses wide-field correction by controlling mostly the low-altitude turbulence, while LTAO uses multiple off-axis WFSs, often using LGSs, to optimise the on-axis correction with a single DM. In all of those cases, the use of a single DM inherently limits the number of degrees of freedom available to the AO system, forcing a compromise between corrected FoV and optical performance. However, with MORFEO, this dichotomy is broken: by adding two post-focal DMs conjugated to different altitudes, it delivers both wide corrected fields and high optical performance, at the cost of increased opto-mechanical complexity, mass, and budget. Figure~\ref{fig:adaptive_optics_zoo} places these modes within the ELT context. 

MORFEO (\url{https://elt.eso.org/instrument/MORFEO/}) is the ELT’s facility MCAO relay, providing uniform, wide-field correction to MICADO and the second gravity-invariant port. It combines wavefront measurements from the six LGSs with up to three NGSs located on the $\sim$160\arcsec\ technical field. It performs tomographic reconstruction to drive two post-focal DMs for high altitudes, and the ELT M4 conjugated to the ground layer. The MCAO architecture corrects for wavefront aberrations and field distortions, providing a nearly diffraction-limited PSF optimised over the MICADO field (51\arcsec), while possibly supporting wider corrected fields of order $\sim$2-2.5\arcmin\ at the second port (\cite{ciliegi_maory_2021}). By decoupling peak performance from field uniformity, MORFEO MCAO enables precision photometry, astrometry, and morphology in fields substantially larger than those otherwise achievable with SCAO alone. MCAO was already proven in the sky by the ESO Multi-Conjugate Adaptive Optics Demonstrator (MAD, (\cite{marchetti_mad_2007})) at the Very Large Telescope and more recently by the GEMS module on the GEMINI South (\cite{neichel_gems_2013}).

\section{MORFEO Architecture and Implementation}\label{sec:morfeo-arch}

\begin{figure}[htbp]
  \centering
  \includegraphics[width=.88\linewidth,height=.34\textheight,keepaspectratio]{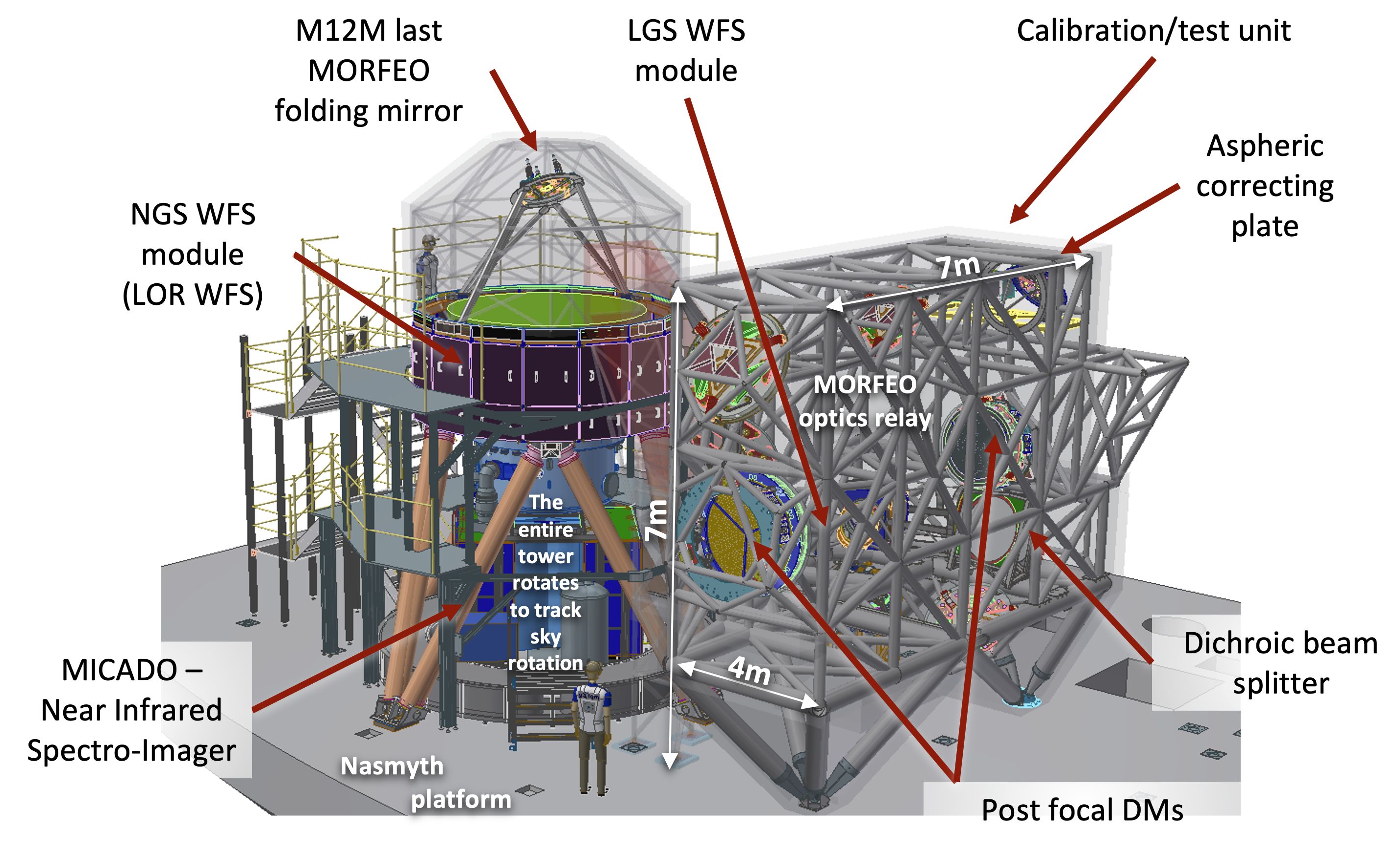}
  \caption{Schematic layout of MORFEO showing the ELT primary, LGS objective, relay optics, and the second-port interface. Credits to Vincenzo Di Caprio.}
  \label{fig:morfeo_schematic}
\end{figure}

MORFEO re-images the ELT focal surface to two gravity-invariant output ports with a 1:1 magnification, while separating laser light with a large dichroic near the pupil and routing it to the LGS wavefront sensors. The main-path optics accommodate two post-focal DMs, which operate in coordination with the telescope adaptive mirror M4 to provide tomographic correction over an arcminute-class field (\cite{ciliegi_maorymorfeo_2022, pariani_morfeo_2022}). See Figure~\ref{fig:morfeo_schematic}.

The optical configuration of MORFEO consists of a reflective relay that re-images the ELT focal surface to two output ports, preserving a 1:1 magnification while minimising wavefront degradation. It employs a refractive correcting plate and four powered mirrors—two spherical and two aspherical—arranged to ensure low wavefront error and geometrical distortion across the corrected field. The first optical element is an aspheric entrance window that improves meta-pupil imaging quality and conjugation of LGS beams, while acting as a thermal and environmental barrier between the telescope and instrument interiors (\cite{magrin_final_2024}). 

Downstream, the optical path includes two large aspherical concave mirrors and two spherical post-focal deformable mirrors: one convex and conjugated at $\sim$17.5\,km, and the other concave and conjugated at $\sim$6.5\,km. This configuration, initially proposed by Delabre and later adapted to MORFEO by Kosmalski and Rakich (\cite{rakich_maxwellian_2020}), achieves excellent image quality and flexibility of alignment within the volume constraints of the Nasmyth platform (\cite{pariani_morfeo_2022}).

System-level tolerance analysis, verified at the Preliminary Design Review, confirmed that the expected RMS wavefront error remains below 100\,nm across the MICADO field of view, with field distortion residuals after calibration remaining within 50\,$\mu$as even under field rotation. The design is robust against thermo-elastic deformations and mechanical misalignments, and allows recovery of performance through periodic fine collimation (\cite{pariani_morfeo_2022,magrin_final_2024}).

In the MCAO scheme, the AO control loop manages at least two DMs based on wavefront measurements in multiple directions. These are means for solving the tomographic wavefront reconstruction (\cite{ragazzoni_adaptive_2000,marchetti_-sky_2007}) of the optical aberrations induced by atmospheric turbulence and the telescope. MORFEO solves the reconstruction problem through the combination of the signals by the Shack-Hartmann (\cite{hartmann_bemerkungen_1900}) wavefront sensor, looking at both the six sodium LGSs (for high-order sensing) and at three NGS (low-order) (\cite{busoni_maorymorfeo_2022}). The MORFEO dichroic transmits the 589 nm LGS light to the LGS WFS module and reflects the near-IR science band and NGS channels to the science ports and NGS WFSs (\cite{bonaglia_morfeo_2024, magrin_final_2024}). The NGS probes pick up the light at the level of the focal plane re-imaged for MICADO, minimising non-common path aberrations. They are mounted on it using XY stages to reach the desired sky position on the so-called green doughnut (\cite{bonaglia_morfeo_2024}).

MORFEO provides two exit ports: one dedicated to the imager-spectrograph, MICADO, (\cite{davies_micado_2021}), and a second gravity-invariant port for HARMONI (\cite{thatte_harmoni_2022}). The MICADO port delivers a corrected field that matches the instrument’s imaging modes (e.g.\ wide field mode, $\sim$51\arcsec\ square at 4\,mas\,px$^{-1}$, or $\sim$20\arcsec\ square at 1.5\,mas\,px$^{-1}$), while the second port was originally conceived for a wider field, of order $\sim$2.0--2.5\arcmin\ in diameter, to enable multiplexed spectroscopy concepts (\cite{ciliegi_morfeo_2024}). 
HARMONI is a diffraction-limited near-infrared integral-field spectrograph covering 0.82--2.45\,$\mu$m, offering four spaxel scales from 60$\times$30\,mas down to 4$\times$4\,mas and spectral resolving powers between $R\!\sim\!3500$ and $R\!\sim\!17000$ (\cite{thatte_harmoni_2022}).

\begin{figure}[htbp]
  \centering
  \includegraphics[width=.88\linewidth,height=.34\textheight,keepaspectratio]{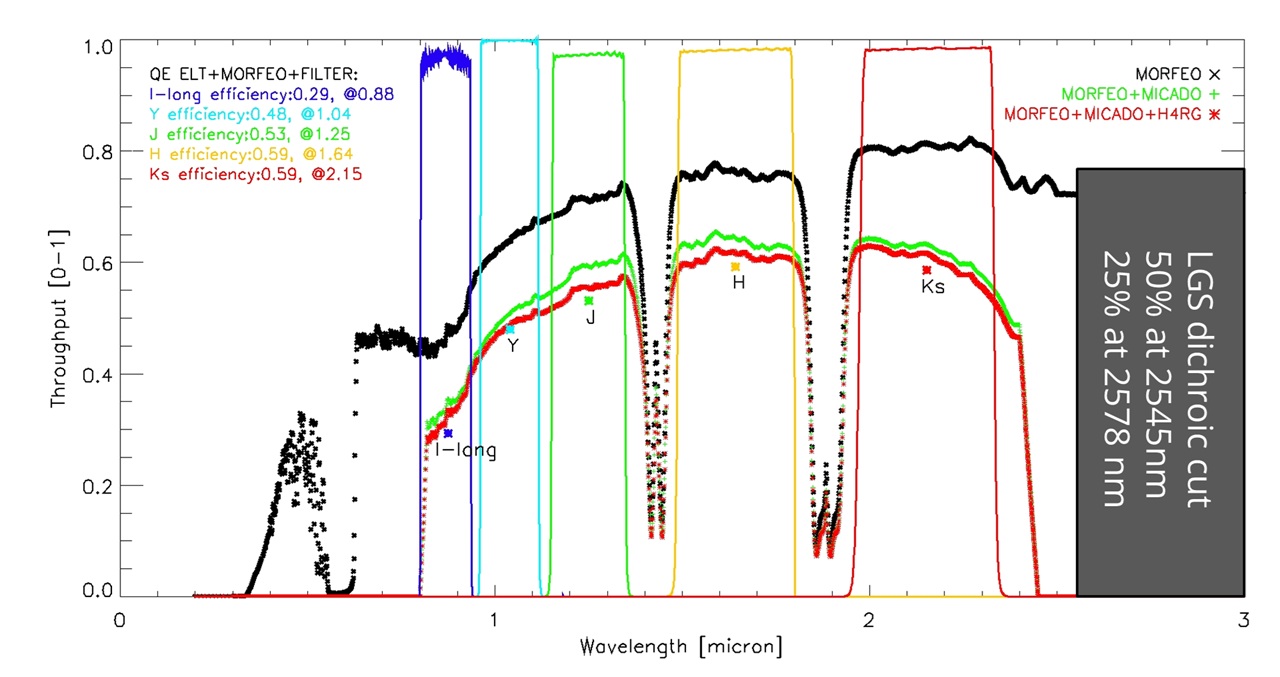}
  \caption{Throughput of the ELT+MORFEO optical train and its combination with MICADO and the H4RG detector, shown together with the instrument passbands. The grey inset highlights the IR cut-off of the LGS dichroic, which transmits the narrow sodium LGS wavelength at 589\, nm while reflecting the science beam over the broad 0.6--2.5\,$\mu$m range. The transition region of the dichroic occurs around 2.545\,$\mu$m (50\% throughput) and 2.578\,$\mu$m (25\%), marking the long-wavelength cut-off of the reflected channel.}

  \label{fig::morfeo_waveband}
\end{figure}

Both exit ports are located downstream of the LGS dichroic (\cite{munari_morfeo_2024, bianco_morfeoelt_2024}), which transmits the 589\,nm laser guide star wavelength while reflecting the science and WFS beams over the wide 600--2500\,nm band toward the MORFEO optical relay, see black crosses in Figure~\ref{fig::morfeo_waveband}.

The MORFEO instrument will be installed on the ELT Nasmyth B platform (\cite{farinato_maorymorfeo_2022}), feeding MICADO via a gravity-invariant interface; a beam steering mirror within MORFEO directs the corrected wavefront to MICADO or to HARMONI (\cite{ciliegi_morfeo_2024}). The instrument's mechanical architecture (\cite{caprio_main_2024}) (rotating tower, co-rotating cabinets and cable-wrap) is designed to respect Nasmyth constraints while maintaining optical alignment and calibration stability.

The real-time computer (RTC) in MORFEO receives the wavefront measurements from both LGS and NGS wavefront sensors, executes tomographic reconstruction, and commands the ELT M4 mirror together with the two MORFEO post-focal deformable mirrors at kilohertz rates (\cite{baruffolo_morfeo_2022, capasso_morfeo_2024}). In addition to the primary high-order and low-order control loops, the RTC controls several auxiliary control loops required for stable MCAO operation, including vibration and jitter compensation, pupil registration and tracking, field derotation, and periodic updates of reconstructor matrices and calibration offsets (\cite{lampitelli_description_2024}). Moreover, it generates and logs a high-rate telemetry stream, recording wavefront slopes, DM commands, residuals, and housekeeping parameters, which are then used for online system performance verification and offline for the MICADO PSF-reconstruction pipeline (\cite{grazian_status_2022}).

\section{Scientific Requirements and Design Drivers}\label{sec:reqs}\label{sec:requirements}

The MORFEO requirements fulfil the needs of the science cases presented in the MAORY/MORFEO White Book (\cite{fiorentino_maory_2019}), and were flowed-down into performance and technical budgets validated at the Preliminary Design Review (\cite{cortecchia_maory_2018,arcidiacono_maory_2020, arcidiacono_preliminary_2022}) and soon, with the Final Design Review scheduled for Q1 2026. In brief, the system shall provide near-diffraction-limited performance for MICADO in the near-IR, with uniform image quality across an arcminute-class field and sufficient sky coverage to enable extragalactic science programmes at high Galactic latitude.
Under median atmospheric conditions, MORFEO is required to deliver a K-band Strehl ratio (2.2\,$\mu$m) of at least 30\%; while in the best conditions, the requirement is to achieve 50\%  (\cite{informationesoorg_morfeo_nodate}). These figures anchor the error budget (fitting and tomographic errors, aliasing, temporal lag, vibrations, non-common-path aberrations) and were confirmed by end-to-end simulations and performance maps produced for design reviews (\cite{agapito_semianalytical_2019}). Uniformity of the PSF across the science field is a key point for most of the scientific cases (\cite{arcidiacono_maory_2020}): the requirement is that PSF metrics (e.g. Strehl Ratio, encircled energy, Full Width at Half Maximum: FWHM) vary slowly with field angle to support precision morphology and photometry (\cite{ciliegi_maorymorfeo_2022}).

Achieving high sky coverage is mandatory for deep extragalactic surveys. The top-level requirement specifies $\gtrsim 50\%$ sky coverage at the South Galactic Pole, enabled by six LGSs for high-order sensing and up to three faint NGSs for low-order and truth sensing (respectively guide stars with $H < 20$ mag and R $< 21.5$). 
The allowable NGS constellation is constrained by both brightness and geometry. Candidate stars should lie outside the MICADO imaging field (vignetting it otherwise), whose size is 20\arcsec\ or 51\arcsec, and within a patrol technical FoV extending to 80\arcsec from the field centre, with a small safety margin added at both the inner and outer boundaries. A minimum mutual separation of $\sim25$\arcsec\     between guide stars avoids mechanical interference. A good indicator for the quality of an NGS constellation is the proximity of the average position to the centre of the field and the spatial distribution, both affecting the tomographic reconstruction. These characteristics, together with the H- and R-band magnitude thresholds, define the operative range of MORFEO and the corresponding sky coverage performance. Although the instrument delivers optimal performance using three bright (H$<17$) well-spaced NGSs, it corrects within the technical specification using two opportunely bright NGSs only, and further expanding the sky coverage (\cite{plantet_sky_2022}) using one, even if losing plate scale control. This loss reduces the jitter suppression by increasing atmospheric tip-tilt anisoplanatism, ultimately blurring the PSF for a Gaussian kernel of the order of some PSF FWHM fraction. This effect is larger for longer exposures and proportional to the distance from the available reference star. In the spectroscopy case, this can reduce the energy within the slit/spaxel.

MORFEO must preserve high optical throughput and wavefront quality over the MICADO observing bands from $I$ through $K$ (0.8--2.4\,$\mu$m), matching MICADO’s imaging and long-slit spectroscopy modes (\cite{clenet_micado_2022}). The dichroic (\cite{bianco_morfeoelt_2024}) and relay optics (\cite{magrin_final_2024}) are specified to minimise chromatic residuals and maximise transmission across this range (\cite{ciliegi_morfeo_2024}).

\section{Performance Characterisation}\label{sec:performance}

\begin{figure}[htbp]
  \centering
  \includegraphics[width=.88\linewidth,height=.34\textheight,keepaspectratio]{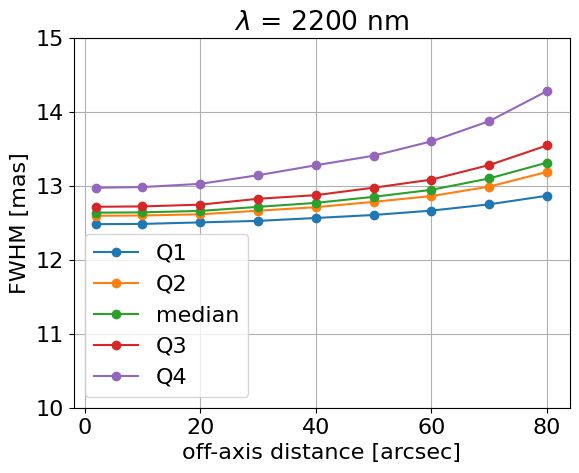}
  \caption{FWHM plot for test stars observed in the Ks filter of MICADO. Using a bright constellation and the different atmospheric conditions, representative of different quartiles once scaled vs performance, median seeing is the one from \cite{sarazin_preliminary_2017}, simulation scaled at 60 degrees elevation. Here, MORFEO used two NGS reference stars of magnitudes H=17 and H=20, corresponding to sky coverage ~40-50\% at the  South Galactic Pole.}

  \label{fig:FWHM}
\end{figure}

\begin{figure}[htbp]
  \centering
  \includegraphics[width=.88\linewidth,height=.34\textheight,keepaspectratio]{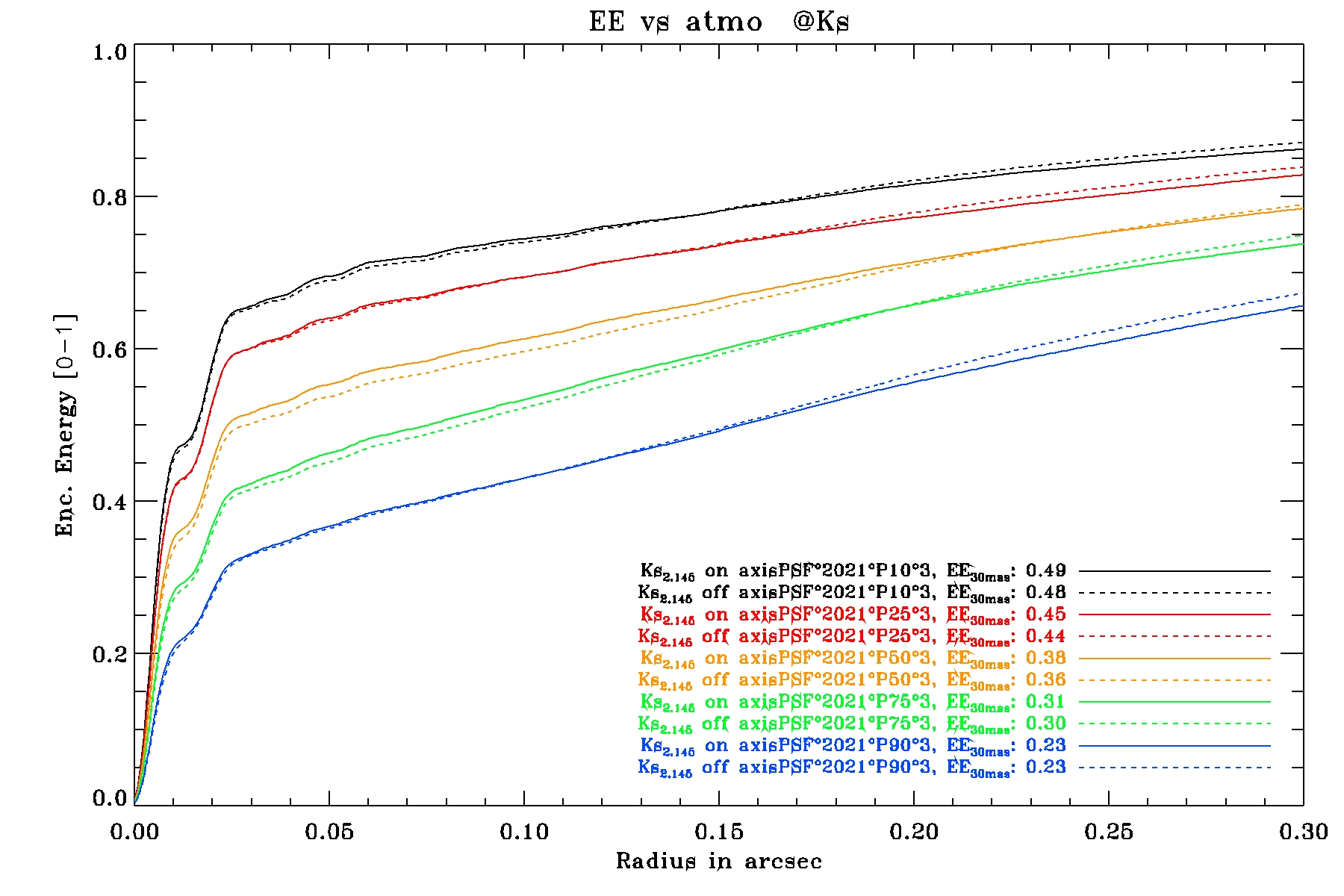}
  \caption{Encircled Energy vs radial distance curves, focusing in the inner 300 mas region from the PSF centre. The PSF is polychromatic and observed in the Ks filter of MICADO. Solid lines represent the centre of the field, while the
dashed line is a direction at the edge (corner) of the 1\arcmin MICADO FoV. The different colours correspond to different atmospheric conditions (10, 25, 50, 75 and 90 percentiles).}

  \label{fig::morfeo_ee}
\end{figure}

End--to--end simulations for MORFEO (\cite{agapito_passata_2016}) predict focal plane at the diffraction limit resolution for most of the cases composing the sky coverage analysis (\cite{plantet_lo_2018}), see Figure~\ref{fig:FWHM}, with median $K$-band Strehl ratios at or above the 30\% requirement with a uniform or at most slowly varying PSF across the MICADO field (\cite{agapito_maory_2020}). Strehl ratio performance (i.e. see Table~\ref{tab:skycoverage_q1_vs_median}) and encircled-energy (EE) curves (i.e. Figure~\ref{fig::morfeo_ee}) over representative guide-star constellations confirm that MCAO provides a much flatter PSF than SCAO over arcminute-class fields (\cite{ellerbroek_first-order_1994}), with EE within fixed apertures remaining stable across the FoV (\cite{busoni_maorymorfeo_2022, ciliegi_maorymorfeo_2022}). In the AO design context, Strehl ratio tracks the PSF core while EE better correlates with photometric/spectroscopic SNR (\cite{arcidiacono_toward_2007}). 

Performance depends on seeing, the structure function of the refractive index vertical profile ($C_n^2(h)$), wind speeds, and sodium-layer properties, as well as on the LGS/NGS asterism and WFS signal-to-noise. Analytical and numerical studies show how tomographic error and tilt anisoplanatism evolve with the turbulence profile, exposure time, and the number/placement of guide stars (\cite{carla_temporal_2022}). Fast PSF predictors (e.g. TIPTOP-style tools (\cite{jolissaint_analytical_2006})) are used to explore these sensitivities efficiently in trade-off analyses and observation planning (\cite{ jolissaint_adaptive_2010,neichel_tiptop_2021}). 

Moreover, on ELT, wind loading on the secondary mirror or the combined action of M4 and M5 during tracking introduces plate-scale variations at the few-milliarcsecond level over arcminute scales on timescales of minutes, with direct consequences for astrometry and encircled energy (\cite{rodeghiero_impact_2018,rodeghiero_performance_2021}). In SCAO, the telescope AO actuators (M4 plus the fast tip--tilt mirror M5) suppress a significant fraction of these perturbations at kilohertz rates, but their corrective capabilities remain limited in an astrometric context. Because M4 is conjugated to the ground layer, it cannot compensate for plate-scale drifts or high altitude-conjugate distortions, and therefore cannot correct geometric distortion as required for high-precision astrometry, since they are intrinsically field-dependent.

This functionality is instead provided by MORFEO: its two post-focal deformable mirrors deliver the field-dependent degrees of freedom required to suppress low-order distortions originating upstream in the telescope optics and MORFEO main path. As shown by recent analyses (\cite{carla_tip-tilt_2022}), the MCAO correction is essential to keep residual telescope-born jitter within the astrometric error budget. The goal of 50--150\,$\mu$ as relative astrometric accuracy across MICADO’s 20\arcsec- zoom mode field can be met with MORFEO, provided adequate calibration and guide-stars constellation are available (\cite{arcidiacono_numerical_2024}).

Furthermore, the MORFEO optical relay has been specifically designed to minimise rotation-induced distortions in the static optics. Since MORFEO remains fixed while the MICADO cryostat rotates for field tracking, residual aberrations would otherwise vary across exposures. However, recent optical modelling shows that these remain within the 50\,$\mu$as distortion budget after calibration, preserving the spatial stability needed for multi-epoch astrometry (\cite{rodeghiero_performance_2021,magrin_final_2024}).

High-rate AO telemetry (WFS slopes, DM commands, reconstructed modes) feeds quality control, performance trending, and PSF reconstruction (PSF-R). For MICADO, a telemetry-only, blind PSF-R service (applicable to both SCAO and MCAO) is being delivered and has passed FDR, with validation on simulated ELT data and on real SCAO datasets from instruments of similar architecture (\cite{grazian_status_2022}). In operations, these products provide per-exposure PSF characterisation (Strehl ratio, FWHM, EE) and enable sensitivity analyses versus atmospheric and system parameters.

An example of the optical performance expressed in terms of EE in 30mas (a good spatial size for a spaxel in a IFU or a slit on the focal plane) is presented in Figure~\ref{fig::morfeo_ee} as a function of the input optical turbulence (defined as a combination of seeing, wind speed and vertical profile and ordered in terms of optical performance after the MORFEO correction). As an example of how the probability to find a given Strehl ratio versus the atmospheric parameters is in Table~\ref{tab:skycoverage_q1_vs_median}.

\begin{table}[htbp]
\centering
\caption{Sky coverage versus Strehl ratio at 2.2\,$\mu$m for two atmospheric conditions: Q1 (0.43\arcsec) and the median value (0.65\arcsec), as in \citet{ciliegi_morfeo_2024}.}
\label{tab:skycoverage_q1_vs_median}
\begin{tabular*}{.7\linewidth}{@{\extracolsep{\fill}}lcc@{}}
\hline
\textbf{SR$_{K}$} & \textbf{Q1, 0.43\arcsec} & \textbf{Median, 0.65\arcsec} \\
\hline
0.10 & 0.97 & 0.95 \\
0.20 & 0.96 & 0.94 \\
0.30 & 0.96 & 0.88 \\
0.40 & 0.93 & 0.66 \\
0.50 & 0.76 & 0.02 \\
0.55 & 0.48 & 0.00   \\
\hline
\end{tabular*}
\end{table}

\section{PSF Reconstruction with MORFEO and MICADO}\label{sec:psfr}

PSF modelling at ELT scale is challenged by temporal variability (servo lag, vibrations), field dependence (anisoplanatism, tomography residuals), and the lack of bright stars in high-Galactic latitude fields, where empirical, star-based PSF maps are unreliable. These issues motivate a telemetry-driven approach so that robust PSFs are available even when the science field is sparse or extended (\cite{simioni_micado_2020,grazian_status_2022}).

MICADO provides a \emph{blind} PSF-Reconstruction (PSF-R) service that uses only AO and facility telemetry (e.g. WFS slopes, DM commands, reconstructor/filters, RTC state) to compute the long-exposure PSF per frame, independent of the science frames. The method is based on traditional control-loop techniques and their adaptations to more recent AO architectures (\cite{veran_estimation_1997,jolissaint_adaptive_2010,arcidiacono_brute_2022}) eventually transposed as Optical Transfer Function (OTF) combination (\cite{wagner_point_2018}). The service is designed to operate across observing modes and wavelengths, to return PSF metrics (Strehl ratio, FWHM, EE) together with the reconstructed PSF image for downstream pipelines (\cite{simioni_micado_2020,grazian_status_2022}).

The MICADO PSF-R will operate for both SCAO and MCAO with MORFEO. In SCAO, the method reconstructs the PSF in the direction of the reference and over the field through temporal tomography (\cite{wagner_direction_2022}). In MCAO, it exploits tomographic telemetry (LGS/NGS WFS streams and multi-DM commands) to deliver a field-dependent PSF (\cite{grazian_status_2022}). This enables homogeneous PSF characterisation for imaging and spectroscopy, also in high galactic latitude fields where few or no suitable stars are present.

Operationally, the PSF-R service retrieves the telemetry synchronous to each science exposure via an archive query, reconstructs the PSF, and hands products to the MICADO pipeline for photometry, morphology, spectroscopic flux calibration, and a posteriori quality control  (\cite{simioni_micado_2020}). The algorithms have been validated on real SCAO data from LBT/SOUL+LUCI, achieving agreement at the $\sim$1--7\% level in Strehl ratio, $\sim$2--5\% in FWHM, and $\sim$4\% in core encircled energy when comparing reconstructed and observed PSFs. Similar tests on ERIS commissioning datasets show comparable accuracy, bringing the technology readiness level to 7 ahead of ELT operations (\cite{simioni_lbt_2022,simioni_point_2022,simioni_towards_2023}).

\section{Observational Tools and Exposure Time Calculators}\label{sec:etc}

Planning observations with MORFEO and its associated instruments requires tools capable of modelling both the telescope-instrument optical train and the AO-dependent PSF. Two complementary approaches are available: the web-based Exposure Time Calculator (ETC) for quick performance estimates and image simulations, and the Python-based \textsc{ScopeSim} ecosystem for physically consistent end-to-end simulations. In parallel, a dedicated ETC is being developed to support the SHARP conceptual design and the exploitation of the MORFEO second port.

\subsection{AETC for MICADO-MORFEO}

The \textsc{Advanced Exposure Time Calculator} (AETC, \url{http://aetc.oapd.inaf.it/}) is a web tool developed by INAF-OAPd to predict imaging performance for various combinations of telescope, instrument and filter (\cite{falomo_aetc_2011}). AETC offers a versatile web-based platform that goes beyond simple exposure‐time estimation. Designed for flexibility, it allows users to define telescope, instrument, filter characteristics and even complex PSF distributions, then simulate the resulting image. Originally configurable for a variety of telescopes and instruments, AETC also supports the configuration of the 39m ELT with MICADO as the client camera. 

In the context of MORFEO, AETC is pre-configured with throughput tables, selectable AO-corrected PSF templates (or user-supplied FITS files), and sky coverage presets. Users can explore how count rates, signal-to-noise ratio, and morphological detectability vary when moving from central to off-axis field positions, or when considering different guide-star geometries. This makes the tool particularly valuable during the early science planning phase, when rapid feasibility assessments, magnitude-limit calculations and field‐layout trials are required.  

Because it handles spatially varying PSFs and image simulation within the same environment, AETC represents an essential first step in bridging the gap between high-level science requirements and detailed observation design for the MORFEO--MICADO system.

\subsection{ScopeSim with MORFEO Extension}

\textsc{ScopeSim} is the Python-based end-to-end simulation environment (\cite{leschinski_scopesim_2020, haberhauer_scopesim_2024}) that allows the propagation of photons through the atmosphere and instrument models defined in an IRDB (Instrument Reference Data Base). When applied to ELT instruments, it creates physically consistent simulations by combining morphology, throughput, background, detector effects, and AO PSFs into one reproducible framework. For MORFEO, a dedicated IRDB package includes ELT surfaces and site conditions, MCAO PSF templates, and MICADO imaging and spectroscopic modes. 

Recent upgrades to \textsc{ScopeSim} now enable support for non-linear spectral traces, integral-field units, and multi-object spectroscopy layouts, improved accuracy in MCAO PSF interpolation across the corrected field and full reproducible workflows via versioned IRDB packages and integration with fast PSF predictors such as \textsc{TIPTOP-iPy} (\cite{neichel_tiptop_2021,haberhauer_scopesim_2024}). Thanks to this possibility, \textsc{ScopeSim} has become the recommended tool for realistic PSF-dependent simulations within the MORFEO system, and it is already adopted by the MICADO and METIS teams for design validation, pipeline development, and performance verification.

\subsection{SHARP-oriented ETC Developments}

A dedicated exposure time calculator (ETC) for the SHARP project is currently being developed, and an early version can be accessed at \url{https://etc-sharp.oas.inaf.it}. The objective of this project is to guide the design and scientific optimisation of the MCAO-assisted spectrograph. In its initial phase, the tool provides signal-to-noise predictions for long-slit and multi-slit modes at diffraction-limited resolution, includes simple parametrised PSFs reflecting MORFEO’s MCAO output, and covers the full corrected 2.5\arcmin\ field. Although the instrument configuration is not yet finalised, this ETC will serve the SHARP science team with a web-based interface for rapid performance assessment and configuration trade-offs.

\section{MORFEO and SHARP}

The MORFEO adaptive optics system has been engineered to accommodate both the primary imager, MICADO, and an additional instrument port, now designated for HARMONI (\cite{ciliegi_maorymorfeo_2022,thatte_harmoni_2022}).
The second port of MORFEO delivers a corrected field of view of approximately 120\arcsec\ in diameter in the baseline design. This wide FoV enables instruments to operate over extended sky areas or exploit high-multiplex configurations, fully benefiting from the uniform MCAO correction provided by MORFEO. The optical performance at the second port is equivalent to that of the MICADO port, since the selection mirror that redirects the beam is located downstream of the last powered optic in the MORFEO relay.

The SHARP multi-object integral field spectrograph (\cite{saracco_sharp_2024,mahmoodzadeh_conceptual_2025}) bases its design on the MORFEO second port and its MCAO features, which can also be applied to other telescopes of similar class equipped with a multi-conjugate adaptive optics module. 
The availability of a wide-field, uniform AO-corrected relay significantly enhances the scientific performance of any and in particular ELT instruments. Thanks to MORFEO, these will push operational modes, also beyond pure imaging, such as high-multiplex, high-spatial-resolution spectroscopy at the highest sensitivity delivered by the telescope. Regarding the MORFEO second port, throughput, field uniformity, and interface stability will be the main strengths of instrument design and operational planning. 

\section{Conclusion}

The \textsc{MORFEO} adaptive-optics relay represents a key technological advancement for the \textsc{ELT}, delivering wide-field, diffraction-limited correction in the near-infrared and enabling a new era of high-precision astronomy. Designed to serve both its primary imager, \textsc{MICADO}, and a future second-port instrument, MORFEO paves the way to science cases that combine high angular resolution with large corrected fields of view. By deploying a constellation of six laser guide stars and up to three natural guide stars in a tomographic reconstruction MCAO scheme, along with multiple deformable mirrors, MORFEO fully meets its technical requirements by delivering levels of correction uniformity and sky coverage, enabling multiple outstanding science drivers.

In particular, the integration of multi-conjugate adaptive optics (MCAO) with telemetry-based point-spread-function reconstruction (PSF-R) equips MORFEO to handle challenging observational regimes: sparse stellar fields, extended emission regions, and high-redshift galaxies, all requiring stable, well-characterised PSFs. The wide FoV and the PSF uniformity are especially critical for spectroscopic follow-up and multiplexed observations, where the knowledge of the optical efficiency across the field is essential. 

The SHARP project will benefit from the MORFEO development, adopting it as a reference optical and AO architecture for its design studies and performance simulations. By relying on MORFEO's validated MCAO framework, SHARP can explore spectroscopic strategies that exploit uniform PSF quality and extended corrected fields. In this way, MORFEO not only supports first-light science at the ELT but also establishes a versatile foundation for the next generation of near-infrared instrumentation.

\section*{Acknowledgement}
The authors thank the entire MORFEO team at INAF and the partner institutions for their contribution to the development, design studies, simulations, and coordination activities. In particular, we acknowledge the efforts of the MORFEO science and engineering groups in advancing the optical and real-time control architecture. 

We also thank the ELT Programme Office at ESO for fruitful collaboration and technical interface support during system design reviews. Simulations and performance studies presented in this work have benefited from computational resources provided by INAF–OAA Arcetri, INAF–OAS Bologna and INAF–Padova.

This research has made use of the AETC exposure time calculator and the ScopeSim simulation environment, developed and maintained by the MICADO and MORFEO instrument teams. The authors acknowledge the use of TIPTOP-iPy tools for PSF modelling and validation.

C.A. acknowledges partial support from the ELT-MORFEO Phase B contract and coordination activities within the ELT instrument consortia.
E.P. was supported by the RSN1 Minigrant of the Istituto Nazionale di Astrofisica (INAF). 

\section*{Declaration of generative AI and AI-assisted technologies in the manuscript preparation process}
During the preparation of this work, the authors used ChatGPT 5.1 to proofread. After using this tool/service, the authors reviewed and edited the content as needed and takes full responsibility for the content of the published article.

\end{document}